\documentclass[twocolumn,showpacs,preprintnumbers,amsmath,amssymb]{revtex4}

\usepackage[dvips]{graphicx}

\begin{document}

\title{Neutrino-induced  coherent pion production off nuclei reexamined}

\author{T.\ Leitner, U.\ Mosel, S.\ Winkelmann\footnote{Present address: Physikalisches Institut, Universit\"at Freiburg}}

\affiliation{Institut f\"ur Theoretische Physik, Universit\"at Giessen, Germany}

\date{January 19, 2009}

\begin{abstract}
  It is pointed out that so far all theoretical estimates of coherent pion production off
  nuclei induced by neutrinos rely on the ``local approximation'' well known in photonuclear
  physics. The effects of dropping this approximation are discussed. It is found that in a
  plane wave approximation for the pion, the local approximation overestimates the coherent
  neutrino-induced pion production on nuclei.
\end{abstract}


\maketitle


A precise determination of neutrino oscillation parameters from long baseline experiments
requires a precise knowledge of the incoming neutrino energy and a detailed understanding
of the neutrino interaction within the detection target. Since the targets in all present
experiments involve heavy nuclei, from C to Fe, a precise knowledge of the interaction
of neutrinos with nuclei is obviously required (for a collection of recent references see Ref.~\cite{Nuint:2007}). In the few GeV region, this involves pion production channels, and a
detailed study of this interaction is necessary to improve the accuracy of any
determination of the neutrino energy spectrum. In this energy regime, the K2K
Collaboration has observed a significant deficit in forward lepton direction when
comparing their measurements with simulations \cite{Aliu:2004sq}.  Since the observed
deficit is present at forward angles, i.e., small momentum transfers, charged current (CC)
coherent pion production is one candidate interaction that could explain the difference. More
recent searches, however, found no evidence for CC coherent pion production in the low-GeV
region \cite{Hiraide:2008eu,Hasegawa:2005td}. The MiniBooNE Collaboration, on the other hand, has reported
neutral current (NC) coherent $\pi^0$ production \cite{AguilarArevalo:2008xs} at neutrino
energies less than 1 GeV.  All these experimental analyses have in common that fact that the
coherent fraction is not accessible directly but has to be extracted from data assuming
specific models for incoherent pion production; the theoretical models for coherent
scattering used in the Monte Carlo event simulations give too large contributions not seen
experimentally \cite{Hiraide:2008eu,Hasegawa:2005td,AguilarArevalo:2008xs}.

This has triggered a large number of theoretical publications on this subject. The
theories generally fall into two categories. In one class of theories, the partially conserverd axial current (PCAC) is used from
the outset, and the coherent neutrino-induced pion production is related to a pion forward
scattering amplitude. This class of models assumes that the process is dominated by the
axial current and that specific nuclear effects play no role besides providing nuclear
size information. A very recent example of this type of approach is the work by Berger
and Sehgal \cite{Berger:2008xs}. In a fully consistent calculation, the models of this
first class should emerge as approximations of the second class of models, which start from
a theoretical description of the nuclear structure and sum the pion production amplitude
coherently over all target nucleon states.  A very recent example of this approach is
given by the work by Amaro et al.\ \cite{Amaro:2008hd}, which also contains a rather
complete list of references to both approaches, the PCAC based and the ones based on
nuclear structure.

All these studies of neutrino-induced reactions are closely related to the
experiments on coherent photon-induced pion and eta production on nuclei
\cite{Koch:1989xf,Rambo:1999jz,Weiss:2001yy,Siodlaczek:2001mh,Krusche:2002iq}.  While the
latter depend only on vector currents, they are, of course, closely related to those
involving neutrinos in the incoming channel. These reactions were often investigated
theoretically by invoking the ``local approximation''. This simplifies the hadronic current
significantly, because it allows one to pull the $\Delta$ propagator out of the pion production
amplitude \cite{Tiator:1984hv}, which becomes local, and as a consequence allows one
to separate out the nuclear form factor. Peters et al.\ \cite{Peters:1998mb,Peters:1998hm}
have discussed the accuracy and the limitations of the local approximation for the case of
photon-induced coherent $\pi^0$ production as well as for coherent $\eta$ production (see
also Refs.~\cite{Tiator:1984hv,AbuRaddad:1997ai,AbuRaddad:1998mq}).

The local approximation is the starting point also used by the theoretical studies of
neutrino-induced coherent nuclear pion production (see Ref.~\cite{Amaro:2008hd} and references 
therein). In all these studies, the pion production amplitude is factorized into a part
that contains the pion production amplitude and one that contains the nuclear size
information. Only via this approximation does the nuclear form factor emerge in the
expressions for coherent pion production on nuclei. Since the results of both the
PCAC-based models and the nuclear physics models rely on this factorization, it is of
interest to investigate how good this approximation actually is for neutrino-induced
processes. In this brief report, we present the results of such an investigation in which we compare the results of a calculation using the local approximation with the results of a
full calculation without this approximation, both without pion final state interactions.

We now briefly discuss the full calculation and then show how the local approximation
leads to the expressions commonly used for calculating the coherent pion production cross
section.

We assume that the pions are dominantly created via the $\Delta(1232)$ resonance; in
Refs.~\cite{AlvarezRuso:2007it,Amaro:2008hd} it was shown that this is a very good
approximation for coherent pion production. The hadronic current for a nucleon is then
given by
\begin{eqnarray} \label{Jhadr}
\lefteqn{J^\mu_{\rm nucleon}(p,q)} \nonumber \\ &=& i
  \frac{f^*}{m_\pi} C^\Delta F(p_\Delta^2)\, \bar{u}(\vec{p}\,') k_\pi^\alpha G_{\alpha
    \beta}(p_\Delta) \Gamma^{\beta \mu}(p,q) u(\vec{p}) ~,
\end{eqnarray}
where $u$-channel diagrams have been neglected. Here $f^*$ is the $N \Delta \pi$ coupling
constant, $F(p_\Delta^2)$ a form factor depending on the invariant mass of the $\Delta$,
$k_\pi$ the pion momentum, $p'$ and $p$ the nucleon's final and initial momenta, and 
$q$ the transferred four-momentum. Correspondingly, the $\Delta$ momentum is given by
$p_\Delta = p + q$. $C^\Delta$ contains isospin factors; it is given by
\begin{equation}
C^\Delta =
\left\{
\begin{array}{ccc}
\text{CC} & p & \sqrt3 \cos \theta_c \\
\text{CC} & n& \sqrt{\frac{1}{3}} \cos \theta_c \\[1mm]
\text{NC} & p, n &\sqrt{\frac{2}{3}} ~.
\end{array}
\right.
\end{equation}
The quantity $G_{\alpha \beta}$ represents the full Rarita-Schwinger (RS) propagator
\begin{equation} \label{Deltaprop}
G_{\alpha \beta} = \frac{1}{p_\Delta^2 - M_\Delta^2 + i M_\Delta \Gamma_\Delta} \, P_{\alpha \beta} ~,
\end{equation}
where $P_{\alpha \beta}$ is the usual RS projection operator. The vertex function
$\Gamma^{\beta \mu}$ in Eq.~(\ref{Jhadr}) denotes the standard electroweak vertex
structure with vector and axial contributions including the resonance excitation form
factors for which we take the same set as applied in Ref.~\cite{AlvarezRuso:2007tt} (for
details see Refs.~\cite{Hernandez:2007qq,Winkelmann-Dipl}). The propagator
[Eq.~(\ref{Deltaprop})] does not include any in-medium changes of the $\Delta$ spectral
function. Their effect is significant, however, they affect both, the local approximation
and the full calculation.  In the target nucleus considered here, $^{12}$C, only $s_{1/2}$
and $p_{3/2}$ states are occupied. The latter reach farther out than the former and have a
higher degeneracy. The effect of in-medium changes of the $\Delta$ spectral function is
thus expected to be smaller in the full calculation than in the one using the local
approximation where the same spectral function is used for all states [cf.\ Eq.\
(\ref{Jlocal})].

The dyadic product of the currents in Eq.~(\ref{Jhadr}) yields the hadronic tensor, which, in
turn, determines the $\nu + N \to \ell + N + \pi$ cross section.

In coherent pion production, the single particle current [Eq.~(\ref{Jhadr})] has to
be summed over all occupied states of the target nucleus. This total current then yields
the hadronic tensor for coherent $\pi$ production on nuclei. This is in contrast to
incoherent processes, where---in the spirit of the impulse approximation---one sums
over the hadronic tensor of each nucleon.

Assuming a mean field model for the nucleons gives immediately a momentum distribution for
the nucleon states. This necessitates the appearance of a momentum integral in the
hadronic current
\begin{equation} \label{Jnucl}
J^\mu_{\rm nucleus}(q) = \sum_{\text{nucleons}} \int d^3\!p \, J^\mu_{\rm nucleon}(p,q)~.
\end{equation}
Here the sum runs over all occupied nucleon single-particle states in the target nucleus.
The single-nucleon current on the right-hand side is obtained from that in
Eq.~(\ref{Jhadr}) by replacing the free-particle spinors $u(\vec{p})$ by the momentum
representations of the bound-state spinors $\psi_i(\vec{p})$ [same for
$\bar{u}(\vec{p}\,')$]. In the following comparison we obtain the latter from a
Walecka-type mean field model for $^{12}$C using parameters from
Ref.~\cite{Peters:1998mb}. We stress that in this most general expression (\ref{Jnucl})
for the nuclear current, the momentum integration extends over the $\Delta$ propagator as
well, since $p_\Delta = p + q$. Because of the presence of the $\Delta$ propagator in the
single-particle currents and the Lorentz structure of the vertex function $\Gamma^{\beta
  \mu}$, coherent pion production does not test the local vector density of the nucleus
but instead the nonlocal structure of various other (tensor) densities. This is the method
used successfully to describe coherent photon-induced $\pi^0$ production on nuclei
\cite{Peters:1998mb}.

The local approximation now consists in fixing the momentum of the initial nucleon state
in the product $G_{\alpha \beta}(p_\Delta) \Gamma^{\beta \mu}(p,q)$ to some value. An
often-used prescription is \cite{Amaro:2008hd}
\begin{equation} \label{statapprox}
\vec{p}\,^0 = -(\vec{q} - \vec{k}_\pi)/2 ~.
\end{equation}
As a consequence of this ``freezing'' of the initial nucleon's momentum in the transition
operator, the momentum of the $\Delta$ resonance is also determined. The propagator of the
$\Delta$ resonance can thus be moved out of the integral and then even out of the sum in
Eq.~(\ref{Jnucl}). This approximation basically consists of suppressing the propagation of
the $\Delta$ resonance and corresponds to the assumption of a very heavy $\Delta$
resonance.  As a consequence, the $W,Z + N \to \pi + N$ vertex becomes local. Thus we are
left with the current in the local approximation
\begin{eqnarray*}
  \tilde{J}^\mu_{\rm nucleus}(q) &=& i \frac{f^*}{m_\pi} C^\Delta F({p^0_\Delta}^2)\, \frac{k_\pi^\alpha}{{p^0_\Delta}^2 - M_\Delta^2 + i M_\Delta \Gamma_\Delta} \,  \nonumber \\
  & & \hspace{-2cm} \times \sum_i \int d^3\!p \,  \left(\psi_i(\vec{p}) \bar{\psi}_i(\vec{p} + \vec{q} - \vec{k}_\pi)\right) P_{\alpha \beta}(p^0_\Delta)
  \Gamma^{\beta \mu}(p^0,q)~,
\end{eqnarray*}
where $i$ runs over all occupied nucleon states and $\vec{p}\,^0_\Delta = (\vec{q} +
\vec{k}_\pi)/2$ according to the local approximation (\ref{statapprox}). In an $r$-space
representation, this becomes
\begin{eqnarray} \label{Jlocal} \tilde{J}^\mu_{\rm nucleus}(q) &=& i \frac{f^*}{m_\pi}
  C^\Delta F({p^0_\Delta}^2)\, \frac{k_\pi^\alpha}{{p^0_\Delta}^2 - M_\Delta^2 + i
    M_\Delta \Gamma_\Delta} \,
  \nonumber \\
  & & \hspace{-2cm} \times \int d^3\!r \, e^{i(\vec{q} - \vec{k}_\pi) \cdot \vec{r}}\,
  {\rm tr} \left( \rho(\vec{r},\vec{r}) P_{\alpha \beta}(p^0_\Delta) \Gamma^{\beta
      \mu}(p^0,q)\right)~.
\end{eqnarray}
Here the trace is taken over the Dirac indices and $\rho(\vec{r},\vec{r})$ is the diagonal
element of the one-body density matrix. This is the final result in the local
approximation.  Equation (\ref{Jlocal}) shows that the nuclear form factor has been factorized
out; all the other (nonlocal) densities present in the full expression no longer appear.

In the following, we compare the full calculation, based on Eq.\ (\ref{Jnucl}) with a
propagating $\Delta$, with the results of the local approximation [Eq.\ (\ref{Jlocal})] for
the target nucleus $^{12}$C. To isolate the effects of the local approximation,
both calculations are done in the plane wave approximation in which the produced pion is
taken to be a free particle. Both calculations use the same nuclear structure model, i.e.,
the density and momentum distributions are calculated consistently in the same
relativistic mean field model. For the local approximation, the incoming nucleon's
momentum has been set to $\vec{p} = 0$, following Ref.~\cite{AlvarezRuso:2007tt}. We note that
when we use the same approximation, we reproduce the local results obtained by Alvarez-Ruso
et al.\ \cite{AlvarezRuso:2007tt}.

\begin{figure}
\includegraphics[width=80mm]{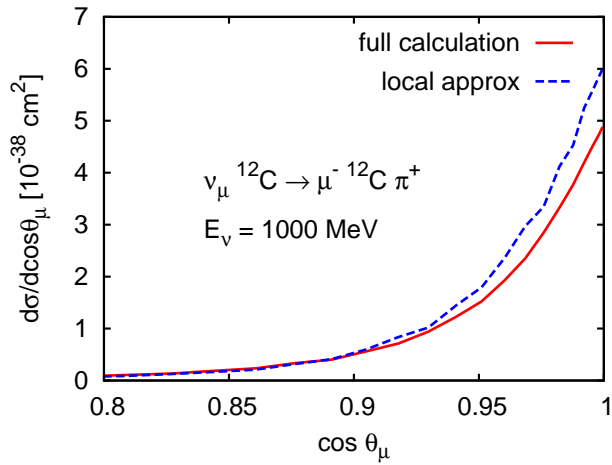}
\caption{\label{fig:dsigmadcos} (Color online) CC induced pion angular distribution for a
  neutrino energy of 1 GeV and target $^{12}$C. The dashed curve gives the result of the
  calculation using the local approximation [cf.\ Eq.\ (\ref{Jlocal})]; the solid curve
  gives the result of a fully dynamic calculation [cf.\ Eq.\ (\ref{Jnucl})]. All curves are
  without pion final state interactions.}
\end{figure}

Figure \ref{fig:dsigmadcos} shows a comparison of the full calculation for the angular
distribution of the produced pions for $E_\nu=1$ GeV with the results obtained by using
the local approximation. Over a wide angular range there is perfect agreement. However, at
very forward angles, the local approximation gives a cross section that is about 20\%
larger than that obtained in the full calculation.

\begin{figure}
\includegraphics[width=85mm]{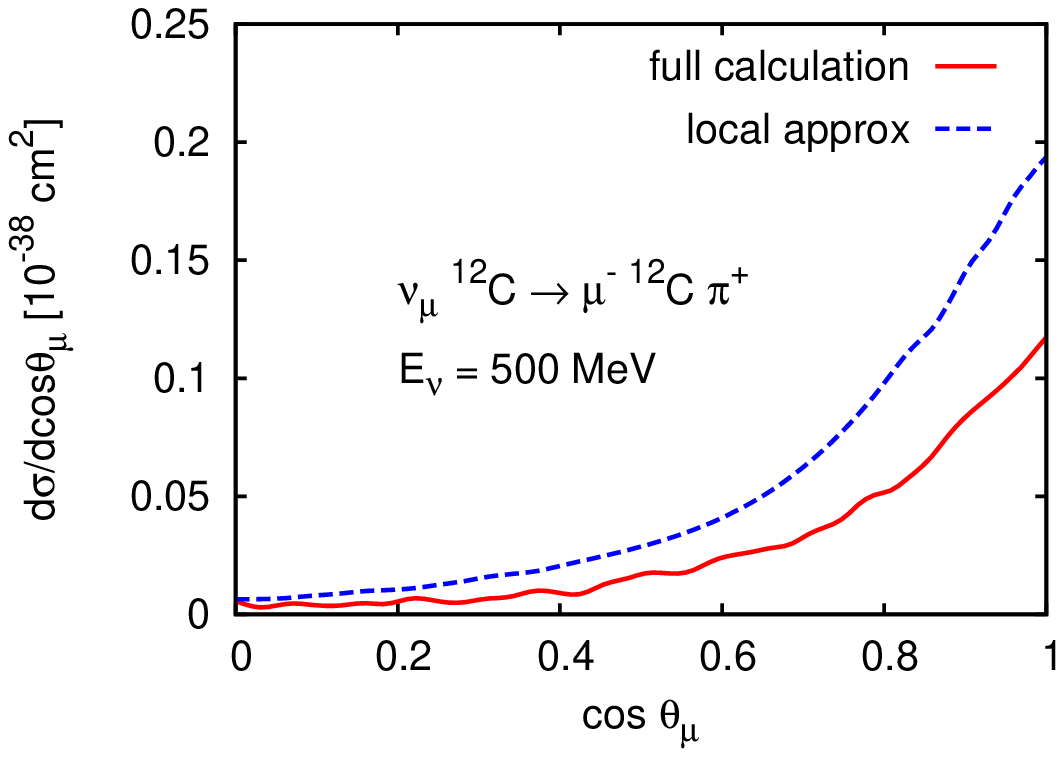}
\caption{\label{fig:dsigmadcos500} (Color online) CC induced pion angular distribution for
  an incoming neutrino energy of 500 MeV and target $^{12}$C. Curves as in
  Fig.~\ref{fig:dsigmadcos}.}
\end{figure}

The difference between the full and the approximate calculation is larger at lower
energies, as can be seen from Fig.\ \ref{fig:dsigmadcos500}, where we compare the full and
the local approximation calculations for the angular distribution at the lower beam energy
of 500 MeV. Here the difference between the two curves is drastic and amounts to a
factor of $\approx 2$ at zero degrees.

\begin{figure}
\includegraphics[width=85mm]{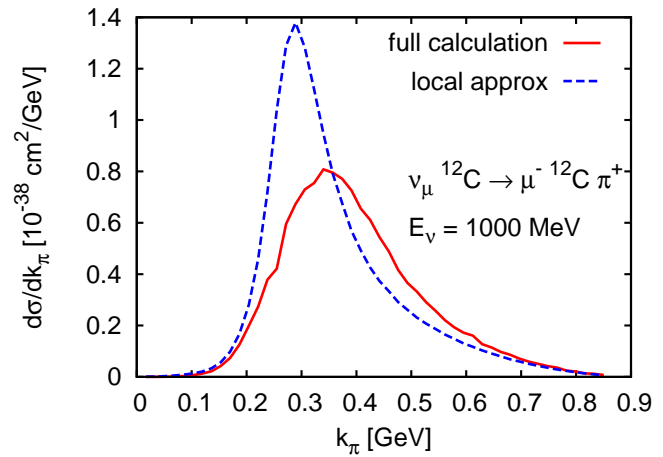}
\caption{\label{fig:dsigmadk} (Color online) CC induced pion momentum distribution for an
  incoming neutrino energy of 1 GeV and target $^{12}C$. Curves as in Fig.\
  \ref{fig:dsigmadcos}.}
\end{figure}

The pion momentum distribution for $E_\nu=1$ GeV is shown in Fig.\ \ref{fig:dsigmadk}. It
is seen that the local approximation overestimates the full result by about 70\% at the
peak. The slight shift downward relative to the fully dynamical result is a consequence of
the local approximation which, as discussed earlier, assumes a very heavy $\Delta$ thus
minimizing any recoil effects. We find qualitatively similar results for NC induced
coherent pion production.
\begin{figure}
\includegraphics[width=85mm]{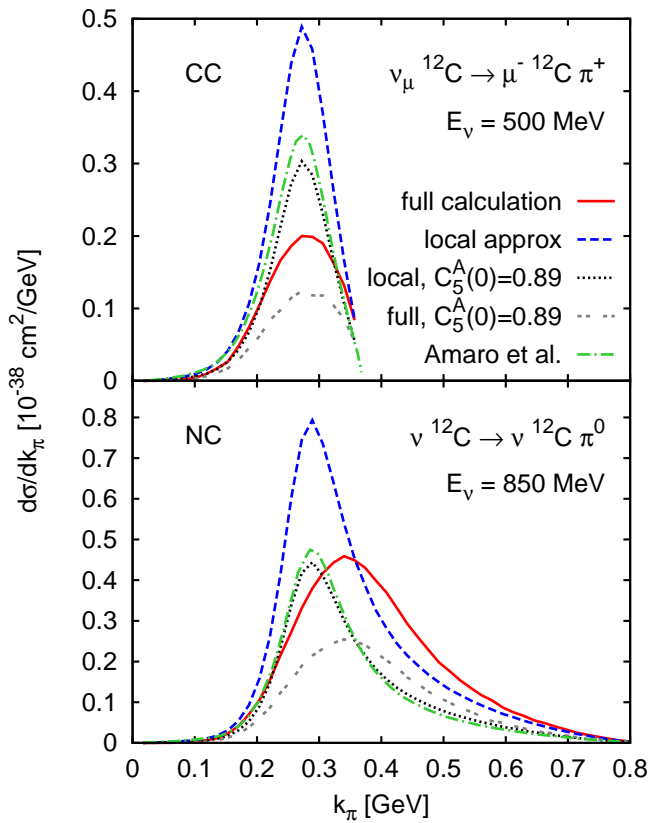}
\caption{\label{fig:dsigmadk1} (Color online) Pion momentum distribution for 500 (CC) and
  850 (NC) MeV incoming neutrino energy. The solid and the dashed curves are as in the
  previous figures, the dotted (double-dashed) ones show the calculation with a modified
  axial coupling for the local approximation (full calculation) as detailed in the text;
  the dash dotted lines are taken from Ref.\ \cite{Amaro:2008hd}.}
\end{figure}

Recently, Amaro et al.\ \cite{Amaro:2008hd} have criticized the assumption of a vanishing
initial nucleon momentum as used in Ref.~\cite{AlvarezRuso:2007tt} (and also in our
calculations for the local approximation). Therefore, in Fig.\ \ref{fig:dsigmadk1}, we
compare the results of our full calculations with the results obtained by Amaro et al.\
\cite{Amaro:2008hd}. The latter calculation does not use an incoming momentum distribution
$\sim \delta(\vec{p})$, but it still employs the local approximation from the start.
Again, the curves shown do not contain any final state interactions of the pion. As
already found above, there is again a drastic disagreement between our full calculation
and our local approximation result (compare the solid and dashed curves). For example, in
the CC case at 500 MeV (upper part of Fig.\ \ref{fig:dsigmadk1}), the full calculation is
less than 1/2 the local approximation result at the peak value.  The result of Amaro et
al.\ (dash-dotted curve), obtained in the local approximation, lies significantly lower
than our ``local'' result.  However, this difference between our local approximation and
their result is mainly due to the use of a significantly reduced axial coupling $C_5^A(0)$
in the work of Amaro et al.\ \cite{Amaro:2008hd}, as can be seen by comparing their result
with the dotted line results in which we also reduced $C_5^A(0)$ from 1.2 to 0.89 in our
calculations.  The small remaining differences between the two local approximation
calculations can be attributed to a somewhat different nuclear structure input and the
different treatment of the incoming nucleon's momentum distribution. The double-dashed
line in Fig.\ \ref{fig:dsigmadk1} shows the full calculation with the reduced value of
$C_5^A$. Again, there is a factor of $\approx 2$ difference between it and the
corresponding local curve (dotted).

So far the plane wave approximation has been used for the outgoing pions, while any
observable cross section contains the effects of the strong pion final state interactions
(FSI). These are known to not only lower the cross section by about 60 \% at the lower
beam energies \cite{Amaro:2008hd}, but also affect the shape of the pion momentum
distributions. In coherent events, the quasielastic pion-nucleon scattering and the
absorption through the $\Delta$ resonance move the peak to lower momenta. We thus expect
that the local approximation result, having more cross section at lower momenta, will be
somewhat more affected by pion FSI than the full calculation result, so that the shapes,
but not the magnitudes, will become closer to each other once pion FSI are included.

In conclusion, all available calculations for neutrino-induced coherent pion production
rely on the local approximation for the elementary interaction vertex. Only this
assumption allows one to factorize out the nuclear form factor. For the case of
photonuclear reactions, this assumption had been scrutinized in many theoretical studies,
starting in the 1980s. For neutrino-induced reactions, however, we do not know of any
previous study of this effect. In the present paper, we show that the use of the local
approximation involves errors in the pion momentum distribution that can reach up to 100\%
and lead to different shapes. The discrepancy decreases with the neutrino energy (more
than a factor of 2 for 500 MeV, about 1.7 for 1 GeV, and less for 2 GeV). The differences
for the pion angular distribution at forward angles also decrease with energy.  They are
of the order of 100\% at 0.5 GeV and 20\% at 1 GeV neutrino energy. There is thus a
general tendency for the local approximation to overestimate the coherent neutrino-induced
pion production.  Full calculations, including pion FSI and medium modifications of the
$\Delta$, which go beyond this brief report, will yield a final clarification of this
point.

\begin{acknowledgments}
  We wish to acknowledge very helpful discussions with Luis Alvarez-Ruso. We are also
  grateful to Stefan Bender for providing us with the momentum-space nuclear wave
  functions. This work has been supported by the Deutsche Forschungsgemeinschaft (DFG).
\end{acknowledgments}


\end{document}